\documentstyle[preprint,tighten,floats,prc,aps]{revtex}

\def\mom{$p_{\rm lab}/A$~=~14.6~GeV/$c$}
\def\reac{Si~+~Au}

\begin{document}

\draft

\title{Low Freeze-out Temperature and High Collective Velocities\\
in Relativistic Heavy-Ion Collisions}

\author{J. Rayford Nix}
\address{Theoretical Division, Los Alamos National Laboratory, Los Alamos, NM
87545, USA}
\date{January 16, 1998}
\maketitle

\begin{abstract}
\hspace{-5.334pt}
On the basis of a nine-parameter expanding source model that
includes special relativity, quantum statistics, resonance decays, and
freeze-out on a realistic hypersurface in spacetime, we analyze in detail
invariant $\pi^+$, $\pi^-$, $K^+$, and $K^-$ one-particle multiplicity
distributions and $\pi^+$ and $K^+$ two-particle correlations in nearly central
collisions of Si~+~Au at $p_{\rm lab}/A$~=~14.6~GeV/$c$.  By considering
separately the one-particle data and the correlation data, we find that the
central baryon density, nuclear temperature, transverse collective velocity,
longitudinal collective velocity, and source velocity are determined primarily
by one-particle multiplicity distributions and that the transverse radius,
longitudinal proper time, width in proper time, and pion incoherence fraction
are determined primarily by two-particle correlations.  By considering
separately the pion data and the kaon data, we find that although the pion
freeze-out occurs somewhat later than the kaon freeze-out, the 99\%
confidence-level error bars associated with the two freeze-outs overlap.  By
constraining the transverse freeze-out to the same source time for all points
with the same longitudinal position and by allowing a more flexible freeze-out
in the longitudinal direction, we find that the precise shape of the freeze-out
hypersurface is relatively unimportant.  By regarding the pion and kaon
one-particle data to be unnormalized, we find that the nuclear temperature
increases slightly, but that its uncertainty increases substantially.  By
including proton one-particle data (which are contaminated by spectator
protons), we find that the nuclear temperature increases slightly.  These
detailed studies confirm our earlier conclusion based on the simultaneous
consideration of the pion and kaon one-particle and correlation data that the
freeze-out temperature is less than 100 MeV and that both the longitudinal and
transverse collective velocities---which are anti-correlated with the
temperature---are substantial.  We also discuss the flaws in several previous
analyses that yielded a much higher freeze-out temperature of approximately 140
MeV for both this reaction and other reactions involving heavier projectiles
and/or higher bombarding energies.
\end{abstract}

\pacs{PACS\@: 25.75.Gz, 25.75.Ld, 21.65.+f, 24.10.Jv}

\section{Introduction}
\label{intro}

Experimentalists around the world are vigorously searching for the quark-gluon
plasma---a predicted new phase of nuclear matter where quarks roam almost freely
throughout the medium instead of being confined to individual nucleons
\cite{Sa,Cs,Wo,qm,qm2}.  Such a plasma is believed to have existed in the first
10~$\mu$s of the universe during the big bang and could be produced in the
laboratory during the little bang in a relativistic heavy-ion collision.

When nuclei collide head-on at relativistic speeds, the nuclear matter is
initially compressed and excited from normal nuclear density and zero
temperature to some maximum values---during which pions, kaons, and other
particles are produced---and then expands, with a decrease in density and
temperature.  The early stages of the process are often treated in terms of
nuclear fluid dynamics, but at some late stage the expanding matter freezes out
into a collection of noninteracting hadrons.

Measurements of the invariant one-particle multiplicity distributions and
two-particle correlations for the pions, kaons, and other particles that are
produced sample the density, temperature, collective velocity, size, and other
properties of the system during this freeze-out.  The use of two-particle
correlations to extract size information was pioneered in the 1950s by
Brown\footnote{Like many individuals throughout history, R. H. Brown preferred
his middle name to his first name.  Therefore, instead of using Robert H. Brown
as author on his many articles and books, he used R. Hanbury Brown instead.
This practice inevitably led to the citation of Brown as Hanbury Brown and even
Hanbury-Brown.  Although this error is widespread throughout the relativistic
heavy-ion community, he is correctly listed under the last name Brown in most
astronomy texts, biographical reference books, and encyclopedias.} and Twiss
\cite{BT}, who used two-photon correlations to measure the size of stars, and by
Goldhaber {\it et al.} \cite{GFGHKP}, who used two-pion correlations to measure
the size of the interaction region in antiproton annihilation.  The hope is that
a sharp discontinuity in the value of one or more of the extracted freeze-out
properties as a function of bombarding energy and/or size of the colliding
nuclei could signal the formation of a quark-gluon plasma or other new physics.

For the extraction of the freeze-out properties from experimental measurements
of invariant one-particle multiplicity distributions and two-particle
correlations, a nine-parameter expanding source model that includes special
relativity, quantum statistics, resonance decays, and freeze-out on a realistic
hypersurface in spacetime was developed in Refs.~\cite{CN,CN2}.  The application
of this model to central collisions of Si~+~Au at $p_{\rm lab}/A$~=~14.6~GeV/$c$
\cite{A+,Ci} led to the conclusion that the freeze-out temperature is less than
100 MeV and that both the longitudinal and transverse collective
velocities---which are anti-correlated with the temperature---are substantial.
Similar conclusions concerning a low freeze-out temperature have also been
reached in Refs.~\cite{CL,CL2}.  However, other analyses
\cite{Xu,B+,BSWX,BSWX2,ECHX} have yielded a much higher freeze-out temperature
of approximately 140 MeV for both this reaction and other reactions involving
heavier projectiles and/or higher bombarding energies.

Because of the importance of resolving this significant discrepancy, we perform
here a more detailed analysis of exactly the same data \cite{A+,Ci} that were
considered in Refs.~\cite{CN,CN2}.  After briefly reviewing the source model in
Sec.~\ref{source} and the results of a simultaneous consideration of the
one-particle data and correlation data in Sec.~\ref{simul}, we examine various
subsets of the data and determine the effect of alternative assumptions on our
results.  In particular, we consider separately the one-particle data and the
correlation data in Sec.~\ref{one-part} and consider separately the pion data
and the kaon data in Sec.~\ref{pion}.  In Sec.~\ref{constr} we constrain the
transverse freeze-out to the same source time for all points with the same
longitudinal position and allow a more generalized freeze-out in the
longitudinal direction, and in Sec.~\ref{unnorm} we regard the pion and kaon
one-particle data to be unnormalized and also include proton one-particle data
(which are contaminated by spectator protons).  These detailed studies provide a
useful background for our discussion in Sec.~\ref{flaws} of the flaws in several
previous analyses that led to an anomalously high freeze-out temperature.  Our
summary and conclusion are given in Sec.~\ref{sum}.

\section{Nine-Parameter Expanding Source Model}
\label{source}

The expanding source model introduced in Refs.~\cite{CN,CN2} describes invariant
one-particle multiplicity distributions and two-particle correlations in nearly
central relativistic heavy-ion collisions in terms of nine parameters, which are
necessary and sufficient to characterize the gross properties of the source
during its freeze-out from a nuclear fluid into a collection of noninteracting,
free-streaming hadrons.  The values of these nine parameters, along with their
uncertainties at 99\% confidence limits, are determined by minimizing $\chi^2$
for the types of data considered.  Several additional physically relevant
quantities, along with their uncertainties at 99\% confidence limits, can then
be directly calculated.  The nine independent source freeze-out properties that
we consider here are the central baryon density $n$, nuclear temperature $T$,
transverse collective velocity $v_{\rm t}$, longitudinal collective velocity
$v_{\ell}$, source velocity $v_{\rm s}$, transverse radius $R_{\rm t}$,
longitudinal proper time $\tau_{\rm f}$, width in proper time $\Delta\tau$, and
pion incoherence fraction $\lambda_{\pi}$.

For a particular type of particle, the invariant one-particle multiplicity
distribution and two-particle correlation function are calculated in terms of a
Wigner distribution function, which is the phase-space density on the freeze-out
hypersurface, giving the probability of producing a particle at spacetime point
$x$ with four-momentum $p$.  It includes both a direct term \cite{BOPSW} and a
term corresponding to 10 resonance decays \cite{Mo}, namely the decay of meson
resonances with masses below 900 MeV and strongly decaying baryon resonances
with masses below 1410 MeV\@.

The direct part of the Wigner distribution function for a particular type of
particle is given by \cite{CN2,BOPSW}
\begin{equation}
\label{e:S}
S_{\!\rm dir}(x,p) = \frac{2J + 1}{(2\pi)^3} \, \frac{p \cdot n(x)}{\exp\{[p
\cdot v(x) - \mu(x)]/T(x)\} \mp 1} ~,
\end{equation}
with the minus sign applying to bosons and the plus sign to fermions.  The
quantity $J$ is the spin of the particle, $v(x)$ is the collective
four-velocity, $T(x)$ is the nuclear temperature, and $\mu(x)$ is the chemical
potential for this type of particle.  The four-vector $n(x)$, with components
\begin{equation}
\label{e:n}
n_\mu(x) = \int_\Sigma d^{3\!}\sigma_{\!\mu}(x')\,\delta^{(4)}(x - x') ~,
\end{equation}
gives the normal-pointing freeze-out hypersurface elements.  The subscript
$\Sigma$ on the integral denotes the limits to the hypersurface for a
finite-sized system.  Because we are considering nearly central collisions, we
assume axial symmetry and work in cylindrical coordinates in the source frame,
with longitudinal distance denoted by $z$, transverse distance denoted by
$\rho$, and time denoted by $t$.  Throughout the paper we use units in which
$\hbar = c = k = 1$, where $\hbar$ is Planck's constant divided by $2\pi$, $c$
is the speed of light, and $k$ is the Boltzmann constant.  However, for clarity,
we reinsert $c$ in the units of quantities whose values are given in the text or
tables.

Integration of Eq.~(\ref{e:S}) over spacetime leads to the Cooper-Frye formula
for the direct contribution to the invariant one-particle multiplicity
distribution \cite{CF}, namely
\begin{equation}
\label{e:P}
P_{\!\rm dir}(p) = E\,\frac{d^{3\!}N_{\rm dir}}{dp^3} = \frac{1}{2\pi m_{\rm
t}}\,\frac{d^{2\!}N_{\rm dir}}{dy\,dm_{\rm t}} = \frac{2J + 1}{(2\pi)^3}
\int_{\Sigma} d^{3\!}\sigma_{\!\mu} \, \frac{p^{\mu}}{\exp\{[p \cdot v(x) -
\mu(x)]/T(x)\} \mp 1} ~,
\end{equation}
where $E$ denotes the particle's energy, $m_t = \sqrt{m^2 + {p_t}^2}$ its
transverse mass, and $y$ its rapidity.  The quantity $m$ is the particle's rest
mass, and $p_t = \sqrt{{p_x}^2 + {p_y}^2}$ is its transverse momentum.  We
assume that the source is boost invariant within the limited region between its
two ends \cite{CFS,Bj}, and that it starts expanding from an infinitesimally
thin disk at time $t = 0$.  The transverse velocity at any point on the
freeze-out hypersurface is assumed to be linear in the transverse
coordinate~$\rho$.

For a particular type of particle, the two-particle correlation function is
given by \cite{CN2,PCZ,CH}
\begin{equation}
\label{e:C}
C(K,q) = \frac{P_2(p_1,p_2)}{P(p_1)\,P(p_2)} = 1 \pm \lambda\,\frac{|\int
d^{4\!}x\, S(x,K)\, \exp(iq\cdot x)|^2}{[\int d^{4\!}x\, S(x,p_1)] [\int
d^{4\!}x\, S(x,p_2)]} ~,
\end{equation}
where $K = \case 1/2 (p_1 + p_2)$ is one-half the pair four-momentum and $q =
p_1 - p_2$ is the pair four-momentum difference.  The plus sign applies to
bosons and the minus sign to fermions, and the quantity $\lambda$ specifies the
fraction of particles of this type that are produced incoherently.

The freeze-out hypersurface is specified by
\begin{equation}
\label{e:tausq}
{\tau_{\rm f}}^2 = \frac{t^2 - z^2}{1 + \alpha_{\rm t}(\rho/R_{\rm t})^2} ~,
\end{equation}
where $\tau_{\rm f}$ is the constant proper time at which freeze-out is assumed
to occur along the symmetry axis of the source and $R_{\rm t}$ is the transverse
radius of the source at the beginning of freeze-out.  The transverse freeze-out
coefficient $\alpha_{\rm t}$ specifies the radial behavior of the freeze-out and
is related to the transverse velocity $v_{\rm t}$ and width $\Delta\tau$ in
proper time during which freeze-out occurs.  We obtain for this relationship
\begin{equation}
\label{e:dtau}
\Delta\tau = \tau_{\rm f} \left[1 - \sqrt{(1 + \alpha_{\rm t})(1 - {v_{\rm
t}}^2)}\right]
\end{equation}
under the additional assumption that the exterior matter at $z = 0$ that freezes
out first has been moving with constant transverse velocity $v_{\rm t}$ from
time $t = 0$ until the beginning of freeze-out.
 
\section{Detailed Analysis of the Reaction \reac \mbox{} at \mom}
\label{analys}

We now use the nine-parameter expanding source model described in
Sec.~\ref{source} to perform a detailed analysis of nearly central collisions in
the reaction Si~+~Au at $p_{\rm lab}/A$~=~14.6~GeV/$c$, for which excellent
experimental data were collected in Experiment E-802 \cite{A+,Ci} at the
Alternating Gradient Synchrotron of the Brookhaven National Laboratory.  We
first describe the results of a simultaneous consideration of the pion and kaon
one-particle and correlation data and then examine various subsets of the data
and determine the effect of alternative assumptions on our results.

\subsection{Simultaneous consideration of pion and kaon one-particle and
correlation data}
\label{simul}

In Experiment E-802, invariant $\pi^+$, $\pi^-$, $K^+$, and $K^-$ one-particle
multiplicity distributions \cite{A+} and $\pi^+$ and $K^+$ two-particle
correlations \cite{Ci} were measured for the central 7\% of collisions in the
reaction Si~+~Au at $p_{\rm lab}/A$~=~14.6~GeV/$c$.  The nine adjustable
parameters of our expanding source model have been determined by minimizing
$\chi^2$ with a total of 1416 data points for the six types of data considered,
so the number of degrees of freedom $\nu$ is 1407.  The error for each point is
calculated as the square root of the sum of the squares of its statistical error
and its systematic error, with a systematic error of 15\% for $\pi^+$, $\pi^-$,
and $K^+$ one-particle multiplicity distributions, 20\% for the $K^-$
one-particle multiplicity distribution, and zero for $\pi^+$ and $K^+$
two-particle correlations \cite{A+,Ci}.  The resulting value of $\chi^2$ is
1484.6, which corresponds to an acceptable value of $\chi^2/\nu = 1.055$.  The
values of nine independent freeze-out properties determined this way, along with
their uncertainties at 99\% confidence limits on all quantities considered
jointly, are given in Table~\ref{t:simul}.  These values and their uncertainties
were obtained earlier in Ref.~\cite{CN2}.  The quantity $n_0$ appearing in
Table~\ref{t:simul} and in the remaining tables denotes normal nuclear density,
whose value is calculated from the nuclear radius constant $r_0$ \cite{MN} by
means of $n_0 = 3/(4\pi{r_0}^3) = 3/[4\pi(1.16\;{\rm fm})^3] = 0.153\;{\rm
fm}^{-3}$.

\begin{table}[h]
\caption{Nine independent source freeze-out properties for central collisions of
Si~+~Au at $p_{\rm lab}/A$~=~14.6~GeV/$c$ resulting from the simultaneous
consideration of the pion and kaon one-particle and correlation data.  The value
used for normal nuclear density $n_0$ is 0.153~fm$^{-3}$.}
\label{t:simul}
\vspace{0.8ex}
\begin{tabular}{lc}

& Value and uncertainty \\
 
Property & \hspace{0pt} at 99\% confidence \\
\hline \vspace{-2.0ex} \\

Central baryon density $n/n_0$ & 0.145 $^{+0.063}_{-0.045}$ \\

Nuclear temperature $T$ (MeV) & 92.9 $\pm$ 4.4 \\

Transverse collective velocity $v_{\rm t}$ ($c$) & 0.683 $\pm$ 0.048 \\

Longitudinal collective velocity $v_{\ell}$ ($c$) & 0.900 $^{+0.023}_{-0.029}$
\\

Source velocity $v_{\rm s}$ ($c$) & 0.875 $^{+0.015}_{-0.016}$ \\

Transverse radius $R_{\rm t}$ (fm) & 8.0 $\pm$ 1.6 \\

Longitudinal proper time $\tau_{\rm f}$ (fm/$c$) & 8.2 $\pm$ 2.2 \\

Width in proper time $\Delta\tau$ (fm/$c$) & 5.9 $^{+4.4}_{-2.6}$ \\

Pion incoherence fraction $\lambda_{\pi}$ & \hspace{0pt} 0.65 $\pm$ 0.11

\end{tabular}
\end{table}

\subsection{Separate consideration of one-particle data and correlation data}
\label{one-part}

We next use our expanding source model to analyze the invariant $\pi^+$,
$\pi^-$, $K^+$, and $K^-$ one-particle multiplicity distributions alone, for
which there are a total of 656 data points.  Because the pion incoherence
fraction $\lambda_{\pi}$ does not enter in the expression for one-particle
multiplicity distributions, there are only eight parameters in this case, so the
number of degrees of freedom $\nu$ is 648.  The resulting value of $\chi^2$ is
692.5, which corresponds to a value of $\chi^2/\nu = 1.069$.  The values of
eight independent freeze-out properties determined this way, along with their
uncertainties at 99\% confidence limits on all quantities considered jointly,
are given in the second column of Table~\ref{t:one-part}.  It is seen that
one-particle multiplicity distributions alone determine the freeze-out central
baryon density, nuclear temperature, transverse collective velocity,
longitudinal collective velocity, and source velocity moderately well, but
contain almost no information concerning the transverse radius, longitudinal
proper time, and width in proper time.

\begin{table}[h]
\caption{Effect on nine independent source freeze-out properties for central
collisions of Si~+~Au at $p_{\rm lab}/A$~=~14.6~GeV/$c$ of considering
separately the one-particle data and the correlation data.  The symbols denoting
the freeze-out properties are defined in Table~\ref{t:simul}.}
\label{t:one-part}
\vspace{0.8ex}
\begin{tabular}{lccc}

& & \multicolumn{2}{c}{\hspace{-10pt}Value and uncertainty at 99\% confidence}
\\

Property & \hspace{72pt} & One-particle data & Correlation data \\
\hline \vspace{-2.0ex} \\

$n/n_0$ & \hspace{72pt} & 0.145 $^{+0.079}_{-0.057}$ & 1.1 $^{+2.5}_{-1.1}$ \\

$T$ (MeV) & \hspace{72pt} & 92.9 $\pm$ 4.7 & 70 $^{+89}_{-70}$ \\

$v_{\rm t}$ ($c$) & \hspace{72pt} & 0.700 $\pm$ 0.095 & 0.95 $^{+0.05}_{-0.36}$
\\

$v_{\ell}$ ($c$) & \hspace{72pt} & 0.904 $^{+0.049}_{-0.094}$ & 0.92
$^{+0.08}_{-0.92}$ \\

$v_{\rm s}$ ($c$) & \hspace{72pt} & 0.872 $^{+0.015}_{-0.017}$ & 0.96
$^{+0.02}_{-0.12}$ \\

$R_{\rm t}$ (fm) & \hspace{72pt} & 9 $^{+20}_{-9}$ & 9.6 $\pm$ 4.9 \\

$\tau_{\rm f}$ (fm/$c$\/) & \hspace{72pt} & 7 $^{+23}_{-7}$ & 9.6 $\pm$ 7.2 \\

$\Delta\tau$ (fm/$c$) & \hspace{72pt} & 3.9 $^{+2.4}_{-3.9}$ & 8.9
$^{+4.3}_{-6.5}$ \\

$\lambda_{\pi}$ & \hspace{72pt} & ---\kern-.02em--- & \hspace{0pt} 0.75 $\pm$
0.18

\end{tabular}
\end{table}

In an analogous study, we use our expanding source model to analyze the $\pi^+$
and $K^+$ two-particle correlations alone, for which there are a total of 760
data points, or 751 degrees of freedom $\nu$.  The resulting value of $\chi^2$
is 756.0, which corresponds to a highly acceptable value of $\chi^2/\nu =
1.007$.  The values of nine independent freeze-out properties determined this
way, along with their uncertainties at 99\% confidence limits on all quantities
considered jointly, are given in the third column of Table~\ref{t:one-part}.  It
is seen that two-particle correlations alone determine the transverse radius,
longitudinal proper time, width in proper time, and pion incoherence fraction
fairly well, but contain almost no information concerning the central baryon
density, nuclear temperature, transverse collective velocity, longitudinal
collective velocity, and source velocity.

\subsection{Separate consideration of pion data and kaon data}
\label{pion}

In our next study, we use our expanding source model to analyze only the data
for pions, namely the invariant $\pi^+$ and $\pi^-$ one-particle multiplicity
distributions and the $\pi^+$ two-particle correlations.  For this case, there
are a total of 934 data points, or 925 degrees of freedom $\nu$.  The resulting
value of $\chi^2$ is 959.9, which corresponds to an acceptable value of
$\chi^2/\nu = 1.038$.  The values of nine independent freeze-out properties
determined this way, along with their uncertainties at 99\% confidence limits on
all quantities considered jointly, are given in the second column of
Table~\ref{t:pion}.  It is seen that when pions alone are considered, the
freeze-out occurs somewhat later and at a lower temperature than when both pions
and kaons are considered (Table~\ref{t:simul}).  However, the 99\%
confidence-level error bars associated with the two freeze-outs overlap.

\begin{table}[h]
\caption{Effect on nine independent source freeze-out properties for central
collisions of Si~+~Au at $p_{\rm lab}/A$~=~14.6~GeV/$c$ of considering
separately the pion data and the kaon data.  The symbols denoting the freeze-out
properties are defined in Table~\ref{t:simul}.}
\label{t:pion}
\vspace{0.8ex}
\begin{tabular}{lccc}

& & \multicolumn{2}{c}{\hspace{-10pt} Value and uncertainty at 99\% confidence}
\\

Property & \hspace{150pt} & Pion data & \hspace{50pt} Kaon data \\
\hline \vspace{-2.0ex} \\

$n/n_0$ & \hspace{150pt} & 0.01 $^{+0.59}_{-0.01}$ & \hspace{50pt} 0.26
$^{+0.17}_{-0.12}$ \\

$T$ (MeV) & \hspace{150pt} & 78 $\pm$ 20 & \hspace{50pt} 101 $\pm$ 10 \\

$v_{\rm t}$ ($c$) & \hspace{150pt} & 0.77 $\pm$ 0.14 & \hspace{50pt} 0.62 $\pm$
0.14 \\

$v_{\ell}$ ($c$) & \hspace{150pt} & 0.920 $^{+0.033}_{-0.056}$ & \hspace{50pt}
0.89 $^{+0.08}_{-0.20}$ \\

$v_{\rm s}$ ($c$) & \hspace{150pt} & 0.880 $^{+0.014}_{-0.015}$ & \hspace{50pt}
0.847 $^{+0.052}_{-0.076}$ \\

$R_{\rm t}$ (fm) & \hspace{150pt} & 9.5 $\pm$ 2.7 & \hspace{50pt} 7.1 $\pm$ 2.5
\\

$\tau_{\rm f}$ (fm/$c$\/) & \hspace{150pt} & 11.1 $\pm$ 5.2 & \hspace{50pt} 6.6
$\pm$ 3.8 \\

$\Delta\tau$ (fm/$c$) & \hspace{150pt} & 8.4 $^{+7.4}_{-4.6}$ & \hspace{50pt}
2.9 $^{+7.4}_{-2.9}$ \\

$\lambda_{\pi}$ & \hspace{150pt} & 0.75 $\pm$ 0.16 & \hspace{50pt}
---\kern-.02em---

\end{tabular}
\end{table}

In an analogous study, we use our expanding source model to analyze only the
data for kaons, namely the invariant $K^+$ and $K^-$ one-particle multiplicity
distributions and the $K^+$ two-particle correlations, for which there are a
total of 482 data points.  Because the pion incoherence fraction $\lambda_{\pi}$
does not enter when kaons alone are considered, there are only eight parameters
in this case, so the number of degrees of freedom $\nu$ is 474.  The resulting
value of $\chi^2$ is 441.5, which corresponds to a highly acceptable value of
$\chi^2/\nu = 0.931$.  The values of eight independent freeze-out properties
determined this way, along with their uncertainties at 99\% confidence limits on
all quantities considered jointly, are given in the third column of
Table~\ref{t:pion}.  It is seen that when kaons alone are considered, the
freeze-out occurs somewhat earlier and at a higher temperature than when both
pions and kaons are considered (Table~\ref{t:simul}).  However, the 99\%
confidence-level error bars associated with these two freeze-outs, as well as
with the separate pion freeze-out and kaon freeze-out, overlap.

\subsection{Constraining and generalizing the freeze-out hypersurface}
\label{constr}

We next consider the effect of constraining the transverse freeze-out to the
same source time for all points with the same longitudinal position, which is
accomplished by holding the transverse freeze-out coefficient $\alpha_{\rm t}$
appearing in Eq.~(\ref{e:tausq}) fixed at zero.  We once again use all 1416 data
points for our six types of pion and kaon one-particle and correlation data.
Because $\alpha_{\rm t}$ is held fixed, there are only eight parameters in this
case, so the number of degrees of freedom $\nu$ is 1408.  The resulting value of
$\chi^2$ is 1519.4, which corresponds to a value of $\chi^2/\nu = 1.079$.  The
values of nine independent freeze-out properties determined this way, along with
their uncertainties at 99\% confidence limits, are given in the second column of
Table~\ref{t:constr}.  Because of the strong dependence of the width in proper
time $\Delta\tau$ upon the transverse freeze-out coefficient $\alpha_{\rm t}$,
which is held fixed at zero, the value of $\Delta\tau$ and its uncertainty are
anomalously small in this case.  We see that constraining the freeze-out
hypersurface in this way affects the remaining eight quantities only to within
their 99\% confidence-level error bars.

\begin{table}[h]
\caption{Effect on nine independent source freeze-out properties for central
collisions of Si~+~Au at $p_{\rm lab}/A$~=~14.6~GeV/$c$ of constraining the
transverse freeze-out to the same source time for all points with the same
longitudinal position (transverse freeze-out coefficient $\alpha_{\rm t}$ fixed
at zero) and of allowing a more flexible freeze-out in the longitudinal
direction (longitudinal freeze-out coefficient $\alpha_{\ell}$ fixed at 0.20
$c^{-2}$).  The symbols denoting the freeze-out properties are defined in
Table~\ref{t:simul}.}
\label{t:constr}
\vspace{0.8ex}
\begin{tabular}{lccc}

& & \multicolumn{2}{c}{\hspace{-18pt}Value and uncertainty at 99\% confidence}
\\

Property & \hspace{72pt} & $\alpha_{\rm t}$ fixed at zero & $\alpha_{\ell}$
fixed at 0.20 $c^{-2}$ \\
\hline \vspace{-2.0ex} \\

$n/n_0$ & \hspace{72pt} & 0.139 $^{+0.058}_{-0.042}$ & 0.141
$^{+0.061}_{-0.043}$ \\

$T$ (MeV) & \hspace{72pt} & 92.6 $\pm$ 4.2 & 92.5 $\pm$ 4.4 \\

$v_{\rm t}$ ($c$) & \hspace{72pt} & 0.721 $\pm$ 0.033 & 0.663 $\pm$ 0.049
\\

$v_{\ell}$ ($c$) & \hspace{72pt} & 0.903 $^{+0.022}_{-0.027}$ & 0.925
$^{+0.025}_{-0.037}$ \\

$v_{\rm s}$ ($c$) & \hspace{72pt} & 0.875 $^{+0.014}_{-0.016}$ & 0.873
$^{+0.014}_{-0.016}$ \\

$R_{\rm t}$ (fm) & \hspace{72pt} & 9.5 $\pm$ 1.2 & 7.7 $\pm$ 1.5 \\

$\tau_{\rm f}$ (fm/$c$\/) & \hspace{72pt} & 6.3 $\pm$ 1.4 & 9.7 $\pm$ 2.2 \\

$\Delta\tau$ (fm/$c$) & \hspace{72pt} & 1.94 $^{+0.52}_{-0.47}$ & 7.3
$^{+4.5}_{-2.7}$ \\

$\lambda_{\pi}$ & \hspace{72pt} & 0.64 $\pm$ 0.11 & \hspace{0pt} 0.57 $\pm$ 0.10

\end{tabular}
\end{table}

In a related study, we consider the effect of allowing a more flexible
freeze-out in the longitudinal direction.  To accomplish this, we generalize
Eq.~(\ref{e:tausq}) to
\begin{equation}
\label{e:tausq2}
{\tau_{\rm f}}^2 = \frac{t^2 - \alpha_{\ell} z^2}{1 + \alpha_{\rm t}(\rho/R_{\rm
t})^2} ~,
\end{equation}
where the longitudinal freeze-out coefficient $\alpha_{\ell}$ allows the
freeze-out along the symmetry axis of the source to occur with a dependence on
longitudinal distance $z$ that is different from that corresponding to a
constant proper time (so that $\tau_{\rm f}$ no longer has this physical
interpretation).  By minimizing $\chi^2$ with a total of 1416 data points for
the six types of data considered, we determined that the minimum in $\chi^2$
occurs at a value of the longitudinal freeze-out coefficient
$\alpha_{\ell}$~=~0.20~$c^{-2}$.  The resulting freeze-out hypersurface is
similar to that obtained in nuclear fluid-dynamical calculations by
Schlei~\cite{Sc}, which was the original motivation for this generalization.  We
then held $\alpha_{\ell}$ fixed at 0.20~$c^{-2}$, so that once again there are
nine parameters and 1407 degrees of freedom.  The resulting value of $\chi^2$ is
1468.5, which corresponds to an acceptable value of $\chi^2/\nu = 1.044$.  The
values of nine independent freeze-out properties determined this way, along with
their uncertainties at 99\% confidence limits, are given in the third column of
Table~\ref{t:constr}.  It is seen that the primary effect of this generalized
freeze-out hypersurface is to increase somewhat the values of $\tau_{\rm f}$ and
$\Delta\tau$, but that the precise shape of the freeze-out hypersurface
remains relatively unimportant.

\subsection{Use of unnormalized one-particle data and inclusion of proton
one-particle data}
\label{unnorm}

Because many analyses have been performed with unnormalized one-particle data
\cite{Xu,B+,BSWX,BSWX2,ECHX}, we now consider the effect of regarding the pion
and kaon one-particle data to be unnormalized.  Once again, we use all 1416 data
points for our six types of pion and kaon one-particle and correlation data.
Because the pion and kaon one-particle normalization constants are allowed to
vary, there are 11 parameters in this case, so the number of degrees of freedom
$\nu$ is 1405.  The resulting value of $\chi^2$ is 1483.2, which corresponds to
a value of $\chi^2/\nu = 1.056$.  The values of nine independent freeze-out
properties determined this way, along with their uncertainties at 99\%
confidence limits, are given in the second column of Table~\ref{t:unnorm}.  It
is seen that the extracted temperature increases slightly, but that its
uncertainty increases substantially.

\begin{table}[h]
\caption{Effect on nine independent source freeze-out properties for central
collisions of Si~+~Au at $p_{\rm lab}/A$~=~14.6~GeV/$c$ of regarding the pion
and kaon one-particle data to be unnormalized and of including proton
one-particle data (which are contaminated by spectator protons).  The symbols
denoting the freeze-out properties are defined in Table~\ref{t:simul}.}
\label{t:unnorm}
\vspace{0.8ex}
\begin{tabular}{lccc}

& & \multicolumn{2}{c}{\hspace{-10pt}Value and uncertainty at 99\% confidence}
\\

& \hspace{72pt} & Unnormalized & Proton one-particle \\

Property & \hspace{72pt} & one-particle data & data included \\
\hline \vspace{-2.0ex} \\

$n/n_0$ & \hspace{72pt} & 0.19 $^{+0.34}_{-0.18}$ & 0.140 $^{+0.039}_{-0.034}$
\\

$T$ (MeV) & \hspace{72pt} & 98 $\pm$ 27 & 97.0 $\pm$ 4.0 \\

$v_{\rm t}$ ($c$) & \hspace{72pt} & 0.65 $\pm$ 0.19 & 0.614 $\pm$ 0.033 \\

$v_{\ell}$ ($c$) & \hspace{72pt} & 0.893 $^{+0.041}_{-0.063}$ & 0.929
$^{+0.016}_{-0.021}$ \\

$v_{\rm s}$ ($c$) & \hspace{72pt} & 0.875 $^{+0.016}_{-0.018}$ & 0.832
$^{+0.020}_{-0.022}$ \\

$R_{\rm t}$ (fm) & \hspace{72pt} & 7.8 $\pm$ 2.4 & 7.4 $\pm$ 1.4 \\

$\tau_{\rm f}$ (fm/$c$\/) & \hspace{72pt} & 8.1 $\pm$ 2.7 & 8.5 $\pm$ 2.0 \\

$\Delta\tau$ (fm/$c$) & \hspace{72pt} & 6.1 $^{+4.7}_{-3.0}$ & 6.6
$^{+3.8}_{-2.7}$ \\

$\lambda_{\pi}$ & \hspace{72pt} & 0.66 $\pm$ 0.15 & \hspace{0pt} 0.66 $\pm$ 0.11

\end{tabular}
\end{table}

Because proton one-particle data are contaminated by the presence of spectator
protons, we have not included them in our analysis thus far.  However, because
they are frequently included in other analyses \cite{Xu,B+,BSWX,BSWX2,ECHX}, we
now consider the effect of including 331 data points for the proton one-particle
multiplicity distribution corresponding to the central 7\% of collisions in the
same reaction that we have been considering \cite{A+}.  A systematic error of
15\% is used also for the proton one-particle multiplicity distribution.  With a
total of 1747 data points for the seven types of data considered and nine
adjustable parameters, the number of degrees of freedom $\nu$ is 1738 in this
case.  The resulting value of $\chi^2$ is 2546.6, which corresponds to an
unacceptably large value of $\chi^2/\nu = 1.465$.  The probability that a
perfect model would have resulted in a value of $\chi^2$ at least as large as
that found here is the incredibly small value $1.1 \times 10^{-33}$.
Nevertheless, for completeness, we give in the third column of
Table~\ref{t:unnorm} the values of nine independent freeze-out properties
determined this way, along with their uncertainties at 99\% confidence limits.

\section{Flaws in Previous Analyses}
\label{flaws}

The results of the above detailed analyses indicate that the freeze-out
temperature is less than 100 MeV and that both the longitudinal and transverse
collective velocities---which are anti-correlated with the temperature---are
substantial.  Similar conclusions concerning a low freeze-out temperature have
also been reached in Refs.~\cite{CL,CL2}.  However, other analyses
\cite{Xu,B+,BSWX,BSWX2,ECHX} have yielded a much higher freeze-out temperature
of approximately 140 MeV for both this reaction and other reactions involving
heavier projectiles and/or higher bombarding energies.  In order to reconcile
this serious discrepancy, we now examine the features in these analyses that
erroneously led them to the conclusion of a much higher freeze-out temperature.
These analyses fall into two major classes, which we consider in turn.

\subsection{Neglect of relativity in extrapolation of slope parameters to zero
particle mass}
\label{neglect}

One type of analysis \cite{Xu,B+} was based upon the extrapolation to zero
particle mass of extracted slope parameters characterizing the dependence of
unnormalized transverse one-particle multiplicity distributions upon transverse
mass.  For a given reaction and type of particle, this transverse one-particle
multiplicity distribution was represented by the expression\footnote{To
facilitate comparisons with our own expressions, we have transformed the
notation used in Refs.~\cite{Xu,B+} to that used here.}
\begin{equation}
\label{e:trans}
\frac{1}{m_{\rm t}}\,\frac{dN}{dm_{\rm t}} = A \exp\!\left(-\frac{m_{\rm
t}}{T_{\rm eff}}\right) ~,
\end{equation}
where $A$ is an arbitrary normalization constant and $T_{\rm eff}$ is the
extracted slope parameter.  Values of $T_{\rm eff}$ were extracted in this way
for six types of particles originating from three separate reactions, namely
$\pi^+$, $\pi^-$, $K^+$, $K^-$, $p$, and $\bar{p}$ originating from the reaction
$p$~+~$p$ at center-of-mass energy $\sqrt s$~=~23~GeV\@, from the 10\% most
central collisions in the reaction S~+~S at $p_{\rm lab}/A$~=~200~GeV/$c$, and
from the 6.4\% most central collisions in the reaction Pb~+~Pb at $p_{\rm
lab}/A$~=~158~GeV/$c$.

As we will see below, the values of these extracted slope parameters contain
valuable information, but they were unfortunately analyzed in Refs.~\cite{Xu,B+}
in terms of the erroneous equation
\begin{equation}
\label{e:erron}
T_{\rm eff} = T + m\bar{v}^2 ~,
\end{equation}
where $T$ is the nuclear temperature (whose value we are trying to determine)
and $\bar{v}$ is the average transverse collective velocity of the expanding
matter from which the particle originated.  Alas, this equation neglects
relativity---even though these are relativistic collisions!  It was introduced
on page 182c of Ref.~\cite{Xu} with the phrase ``One may empirically guess a
relationship between the slope parameter and particle mass,'' whereas the words
describing the same equation on page 2082 of Ref.~\cite{B+} are ``The
correlation between the slope parameter and particle mass $m$ may be described
qualitatively by the relationship \ldots'' On the basis of the erroneous
Eq.~(\ref{e:erron}), the extrapolation in Ref.~\cite{B+} of the extracted slope
parameters to zero particle mass yielded the result
$T$~$\approx$~140~$\pm$~15~MeV\@.

In the limit in which the particle velocity is large compared to the average
collective velocity and with the aid of other simplifying assumptions and
approximations,\footnote{The expanding source model developed in
Refs.~\cite{CN,CN2} and used here does {\it not\/} require that the particle
velocity be large compared to the average collective velocity and does {\it
not\/} utilize these other simplifying assumptions and approximations.} the
correct relationship between slope parameter, nuclear temperature, particle
mass, and average collective velocity can be easily derived from the
relativistically correct Eq.~(\ref{e:P}).  With the neglect of contributions
from resonance decays, the neglect of the $\mp 1$ appearing in the denominator
of Eq.~(\ref{e:P}), the assumption of a constant freeze-out temperature, and the
assumption that freeze-out occurs at a constant time $t$ in the source frame,
Eq.~(\ref{e:P}) leads to
\begin{equation}
\label{e:trans2}
\frac{1}{m_{\rm t}}\,\frac{d^{2\!}N}{dy\,dm_{\rm t}} = A' E \int_V d^{3\!}x 
\exp\!\left[-\frac{p \cdot v({\bf x})}{T}\right] = A' E \int_V d^{3\!}x
\exp\!\left\{-\frac{\gamma({\bf x})[E - {\bf p} {\bf \cdot} {\bf v}({\bf
x})]}{T}\right\} ~,
\end{equation}
where $A'$ is a different arbitrary normalization constant from the one
appearing in Eq.~(\ref{e:trans}), the subscript $V$ on the integral denotes the
spatial integration limits for the source, and the position-dependent Lorentz
factor $\gamma({\bf x}) = 1/\sqrt{1 - {\bf v}({\bf x}) {\bf \cdot} {\bf v}({\bf
x})}$.

By introducing an average collective velocity $\bar{v}$ in the integrations in
Eq.~(\ref{e:trans2}), taking the limit in which the particle velocity is large
compared to the collective velocity, specializing to the transverse direction,
and neglecting the pre-exponential $E$ dependence, we are led to
\begin{equation}
\label{e:trans3}
\frac{1}{m_{\rm t}}\,\frac{dN}{dm_{\rm t}} = A
\exp\!\left[-\frac{\bar{\gamma}(m_{\rm t} - p_{\rm t}\bar{v})}{T}\right] = A
\exp\!\left(-\frac{m_{\rm t} - \bar{v}\sqrt{{m_{\rm t}}^2 - m^2}}{T\sqrt{1 -
\bar{v}^2}}\right) ~.
\end{equation}
To obtain the relationship between the slope parameter, nuclear temperature,
particle mass, and average collective velocity, we equate the derivatives with
respect to $m_{\rm t}$ of Eqs.~(\ref{e:trans}) and (\ref{e:trans3}), which leads
to
\begin{equation}
\label{e:T}
T = \left(1 - \frac{\bar{v}m_{\rm t}}{p_{\rm t}}\right) \frac{T_{\rm
eff}}{\sqrt{1 - \bar{v}^2}} = \left(1 - \bar{v}\sqrt{1 + \frac{m^2}{{p_{\rm
t}}^2}}\right) \frac{T_{\rm eff}}{\sqrt{1 - \bar{v}^2}} ~.
\end{equation}
An analogous relationship has also been obtained by Siemens and Rasmussen
\cite{SR} for the case of a blast wave produced by the explosion of a
spherically symmetric fireball.

In the limit of zero particle mass, Eq.~(\ref{e:T}) reduces to
\begin{equation}
\label{e:T2}
T = T_{\rm eff} \sqrt{\frac{1 - \bar{v}}{1 + \bar{v}}} ~,
\end{equation}
which agrees with the result obtained by Schnedermann, Sollfrank, and Heinz
\cite{SSH,SSH2} for the case of cylindrical symmetry.  With a typical value of
0.4~$c$ for the average collective velocity $\bar{v}$ and the limiting value of
$T_{\rm eff}$~$\approx$~140~$\pm$~15~MeV obtained in Ref.~\cite{B+} by
extrapolating slope parameters to zero particle mass, Eq.~(\ref{e:T2}) yields
$T$~$\approx$~92~$\pm$~10~MeV for the nuclear temperature.

\subsection{Accumulation of effects from several approximations made in a
thermal model}
\label{accum}

Another type of analysis \cite{Xu,B+,BSWX,BSWX2,ECHX} utilized the thermal model
of Schnedermann, Sollfrank, and Heinz \cite{SSH,SSH2} to extract the nuclear
temperature and transverse surface collective velocity from unnormalized
experimental transverse one-particle multiplicity distributions.  An
accumulation of effects from several approximations led to a somewhat higher
temperature than we have found in our expanding source model \cite{CN,CN2}.
These approximations include the neglect of contributions from resonance decays,
the neglect of the $\mp 1$ appearing in the denominator of Eq.~(\ref{e:P}), the
neglect of the coupling of the transverse motion to the longitudinal motion,
and---most importantly---the neglect of information contained in the absolute
normalization of the multiplicity distributions.  The accumulation of effects
from these approximations was responsible for the conclusion on page 2083 of
Ref.~\cite{B+} that ``Within a temperature range 100~$\le$~$T$~$\le$~150~MeV\@,
the fits are equally good.''  Clearly, the use of unnormalized experimental
transverse one-particle multiplicity distributions in such a thermal model
cannot be expected to provide a definitive determination of the nuclear
temperature at freeze-out.

\section{Summary and Conclusion}
\label{sum}

We have used a nine-parameter expanding source model that includes special
relativity, quantum statistics, resonance decays, and freeze-out on a realistic
hypersurface in spacetime to analyze in detail invariant $\pi^+$, $\pi^-$,
$K^+$, and $K^-$ one-particle multiplicity distributions and $\pi^+$ and $K^+$
two-particle correlations in nearly central collisions of Si~+~Au at $p_{\rm
lab}/A$~=~14.6~GeV/$c$.  By considering separately the one-particle data and the
correlation data, we found that the central baryon density, nuclear temperature,
transverse collective velocity, longitudinal collective velocity, and source
velocity are determined primarily by one-particle multiplicity distributions and
that the transverse radius, longitudinal proper time, width in proper time, and
pion incoherence fraction are determined primarily by two-particle correlations.
By considering separately the pion data and the kaon data, we found that
although the pion freeze-out occurs somewhat later than the kaon freeze-out, the
99\% confidence-level error bars associated with the two freeze-outs overlap.

By constraining the transverse freeze-out to the same source time for all points
with the same longitudinal position and by allowing a more flexible freeze-out
in the longitudinal direction, we found that the precise shape of the freeze-out
hypersurface is relatively unimportant.  By regarding the pion and kaon
one-particle data to be unnormalized, we found that the nuclear temperature
increases slightly, but that its uncertainty increases substantially.  By
including proton one-particle data (which are contaminated by spectator
protons), we found that the nuclear temperature increases slightly.  These
detailed studies confirm our earlier conclusion \cite{CN,CN2} based on the
simultaneous consideration of the pion and kaon one-particle and correlation
data that the freeze-out temperature is less than 100 MeV and that both the
longitudinal and transverse collective velocities---which are anti-correlated
with the temperature---are substantial.

We also discussed the flaws in previous analyses that yielded a much higher
freeze-out temperature of approximately 140 MeV for both this reaction and other
reactions involving heavier projectiles and/or higher bombarding energies.  One
type of analysis was based upon the use of an erroneous equation that neglects
relativity to extrapolate slope parameters to zero particle mass.  Another type
of analysis utilized a thermal model in which there was an accumulation of
effects from several approximations.

The future should witness the arrival of much new data on invariant one-particle
multiplicity distributions and two-particle correlations as functions of
bombarding energy and/or size of the colliding nuclei.  The proper analysis of
these data in terms of a realistic model could yield accurate values for the
density, temperature, collective velocity, size, and other properties of the
expanding matter as it freezes out into a collection of noninteracting hadrons.
A sharp discontinuity in the value of one or more of these properties could
conceivably be the long-awaited signal for the formation of a quark-gluon plasma
or other new physics.
  
\acknowledgments 

I am grateful to Scott Chapman for his valuable contributions during the early
phases of this work (especially his development of the robust computer program
{\tt freezer\/} that was used in the present detailed analyses), to Bernd
R. Schlei for illuminating discussions about freeze-out hypersurfaces, to Hubert
W. van Hecke, John P. Sullivan, and Nu Xu for stressing to me that prior
slope-parameter analyses have yielded high freeze-out temperatures, and to
T. Vincent A. Cianciolo for permitting the use of his preliminary data on
two-particle correlations.  This work was supported by the U. S. Department of
Energy.


\begin{references}

\bibitem{Sa}
H. Satz, Annu.\ Rev.\ Nucl.\ Part.\ Sci.\ {\bf 35}, 245 (1985).

\bibitem{Cs}
L. P. Csernai, {\it Introduction to Relativistic Heavy Ion Collisions\/} (Wiley,
Chichester, 1994).

\bibitem{Wo}
C. Y. Wong, {\it Introduction to High-Energy Heavy-Ion Collisions\/} (World
Scientific, Singapore, 1994).

\bibitem{qm}
{\it Quark Matter '96, Proc.\ Twelth Int.\ Conf.\ on Ultra-Relativistic
Nucleus-Nucleus Collisions, Heidelberg, Germany, 1996}, Nucl.\ Phys.\ A {\bf
610}, 1c (1996).

\bibitem{qm2}
{\it Quark Matter '97, Proc.\ Thirteenth Int.\ Conf.\ on Ultra-Relativistic
Nucleus-Nucleus Collisions, Tsukuba, Japan, 1997\/} (to be published).

\bibitem{BT}
R. H. Brown and R. Q. Twiss, Phil.\ Mag.\ {\bf 45}, 663 (1954).  

\bibitem{GFGHKP}
G. Goldhaber, W. B. Fowler, S. Goldhaber, T. F. Hoang, T. E. Kalogeropoulos, and
W. M. Powell, Phys.\ Rev.\ Lett.\ {\bf 3}, 181 (1959).

\bibitem{CN}
S. Chapman and J. R. Nix, in {\it Advances in Nuclear Dynamics 2, Proc.\ 12th
Winter Workshop on Nuclear Dynamics, Snowbird, Utah, 1996\/} (Plenum Press, New
York, 1996), p.\ 7.

\bibitem{CN2}
S. Chapman and J. R. Nix, Phys.\ Rev.\ C {\bf 54}, 866 (1996).

\bibitem{A+}
E-802 Collaboration, T. Abbott {\it et al.}, Phys. Rev. C {\bf 50}, 1024 (1994).

\bibitem{Ci}
E-802 Collaboration, T. V. A. Cianciolo (private communication).

\bibitem{CL}
T. Cs\"org\H{o} and B. L\"orstad, Nucl.\ Phys.\ A {\bf 590}, 465c (1995).

\bibitem{CL2}
T. Cs\"org\H{o} and B. L\"orstad, Acta Phys.\ Hung.\ New Series, Heavy Ion
Physics {\bf 4}, 221 (1996).

\bibitem{Xu}
N. Xu, for the NA44 Collaboration, I. G. Bearden {\it et al.}, Nucl.\ Phys.\ A
{\bf 610}, 175c (1996).

\bibitem{B+}
NA44 Collaboration, I. G. Bearden {\it et al.}, Phys.\ Rev.\ Lett.\ {\bf 78},
2080 (1997).

\bibitem{BSWX}
P. Braun-Munzinger, J. Stachel, J. P. Wessels, and N. Xu, Phys.\ Lett.\ B {\bf
344}, 43 (1995).

\bibitem{BSWX2}
P. Braun-Munzinger, J. Stachel, J. P. Wessels, and N. Xu, Phys.\ Lett.\ B {\bf
365}, 1 (1996).

\bibitem{ECHX}
S. Esumi, S. Chapman, H. van Hecke, and N. Xu, Phys.\ Rev.\ C {\bf 55}, R2163
(1997).

\bibitem{BOPSW}
J. Bolz, U. Ornik, M. Pl\"umer, B. R. Schlei, and R. M. Weiner, Phys.\ Lett.\ B
{\bf 300}, 404 (1993).

\bibitem{Mo}
Particle Data Group, L. Montanet {\it et al.}, Phys.\ Rev.\ D {\bf 50}, 1173
(1994).

\bibitem{CF}
F. Cooper and G. Frye, Phys.\ Rev.\ D {\bf 10}, 186 (1974).

\bibitem{CFS}
F. Cooper, G. Frye, and E. Schonberg, Phys.\ Rev.\ D {\bf 11}, 192 (1975).

\bibitem{Bj}
J. D. Bjorken, Phys.\ Rev.\ D {\bf 27}, 140 (1983).

\bibitem{PCZ} 
S. Pratt, T. Cs\"org\H{o}, and J. Zim\'anyi, Phys.\ Rev.\ C {\bf 42}, 2646
(1990).

\bibitem{CH} 
S. Chapman and U. Heinz, Phys.\ Lett.\ B {\bf 340}, 250 (1994).

\bibitem{MN} 
P. M\"oller and J. R. Nix, Nucl.\ Phys.\ A {\bf 361}, 117 (1981).

\bibitem{Sc}
B. R. Schlei, Acta Phys.\ Hung.\ New Series, Heavy Ion Phys.\ {\bf 5}, 403
(1997).

\bibitem{SR}
P. J. Siemens and J. O. Rasmussen, Phys.\ Rev.\ Lett.\ {\bf 42}, 880 (1979).

\bibitem{SSH}
E. Schnedermann, J. Sollfrank, and U. Heinz, in {\it Particle Production in
Highly Excited Matter, Proc.\ NATO Advanced Study Institute on Particle
Production in Highly Excited Matter, Il Ciocco, Tuscany, Italy, 1992\/} (Plenum
Press, New York, 1993), p.\ 175.

\bibitem{SSH2}
E. Schnedermann, J. Sollfrank, and U. Heinz, Phys.\ Rev.\ C {\bf 48}, 2462
(1993).

\end{references}
\end{document}